\documentclass[a4paper]{article}
\usepackage[T2A]{fontenc}
\usepackage[cp1251]{inputenc}
\usepackage[russian,english]{babel}
\usepackage{geometry}
\usepackage[dvips]{graphicx}
\usepackage{amsmath}
\usepackage{amssymb}
\usepackage{indentfirst}
\usepackage{multicol}
\usepackage{multirow}
\usepackage{longtable}
\usepackage{setspace}

\begin{document}

\renewcommand{\rmdefault}{ftm}
%\renewcommand{\contentsname}{\center ОГЛАВЛЕНИЕ}
%\renewcommand{\refname}{\center СПИСОК ЛИТЕРАТУРЫ}

%\normalsize
%\begin{spacing}{1} \tableofcontents
%\end{spacing}
\title{First-principles calculations atomic structure and elastic
properties of Ti-Nb alloys}
\author{A.N. Timoshevskii, S.О. Yablonovskii\\ Institute of Magnetism, National Academy of Science
Ukraine,\\ 03142 Kiev,Ukraine,\\\\
  O. M. Ivasishin,   \\Institute for Metal Physics National Academy of Sciences,\\
36 Vernadsky str., 03142 Kiev, Ukraine
 }

 \maketitle
 \abstract{Elastic properties of Ti based $\beta$-alloy were studied
by the method of the model structure first principle calculations.
Concentrational dependence of Young modulus for the binary
$\beta$-alloy Ti$_{1-x}$Nb$_x$  was discovered. It is shown that
peculiarities visible at 0.15-0.18$\%$ a.u. concentrations can be
related to the different Nb atoms distribution. Detailed comparison
of the calculation results with the measurement results was done.
Young modulus for the set of the ordered structures with different
Nb atoms location, which simulate triple $\beta$-alloys Ti$_{0.518}$
Zr$_{0.297}$Nb$_{0.185}$ and Ti$_{0.297}$Zr$_{0.518}$Nb$_{0.185}$
have been calculated. The results of these calculations allowed us
to suggest the concentration region for single-phase ternary
$\beta$-phase alloys possessing low values of Young's modulus.}
%\clearpage
%\section{Упругие свойства системы Ti-Nb}\label{elastic_properties}
\fontsize{12pt}{21pt}\selectfont
\renewcommand{\baselinestretch}{1.5}
\section*{Introduction}

Today the $\beta$-titanium alloys are widely used for medical
applications due to their high bio-compatibility and untoxicity. The
elastic moduli of these alloys have to be rather low. Elastic
properties of the $\beta$-alloys of Ti$_{1-x}$Nb$_x$ system have
been studied extensively. It is known that at Nb content below х=0.27
these alloys are not in a single-phase condition and contain
martensite phases \cite{AbdelHady2006477}. At higher Nb content
single-phase conditions can be fixed. Martensite phases usually lead
to a considerable increase in elastic constants of these alloys.
Elastic moduli of the single $\beta$-phase apparently decrease with
lowering Nb content, as the $\beta$-phase turns less stable. Alloys
with low elastic moduli of about 30-40 GPa are promising materials
for implants. So, single-phase $\beta$-alloys with Nb content below
0.27 could be such material, however, as mentioned above they are
practically impossible to obtain, and there is no experimental data
on the $\beta$-phase elastic moduli dependence on Nb content in
Ti$_{1-x}$Nb$_x$ alloys at x<0.27. It is worth to note that knowing this
dependence is an important issue since it concerns changes in the
elastic moduli of binary Ti$_{1-x}$Nb$_x$ alloy when Ti atoms are
partially substituted by other atoms. In this regard single-phase
Ti$_{1-x-y}$Zr$_y$Nb$_x$  ternary alloys with low elastic moduli are
of interest. In contrast to Ti$_{1-x}$Nb$_x$ alloys, the ternary
alloys can be fixed in a single-phase condition at considerably
lower Nb contents (up to х=0.14) \cite{tagkey2007259}. It is well
known also that Ti$_{1-x-y}$Zr$_y$Nb$_x$ alloys have lower elastic
moduli as compared to Ti$_{1-x}$Nb$_x$ alloys. Therefore, partial
substitution of Ti by Zr in Ti$_{1-x}$Nb$_x$ alloy is expected to
have a double effect - it will lower Nb content necessary to fix a
single-phase state, and it will reduce the elastic moduli of the
ternary alloy. At the first stage of search for
Ti$_{1-x-y}$Zr$_y$Nb$_x$ alloy optimal compositions, the information
on the elastic moduli of binary Ti$_{1-x}$Nb$_x$ $\beta$-alloy in a
wide range of Nb content should be obtained. Such information is
valuable also for low Nb contents at which the alloy can not be
practically fixed in a single-phase condition. In this regard
first-principles calculations of electronic structure and elastic
properties of the alloy are useful. High-precision quantum-mechanical
calculations allow to study in detail various factors affecting the
elastic moduli of a material. Apparently the most important factors
are Nb content and distribution of Nb atoms in the BCC lattice.
Whether the elastic moduli dependence on Nb content represents a monotonic
function or not is also a question of interest. The probability of
an abnormal behavior of this dependence at some Nb contents should
not be excluded. If so, it would be useful to elucidate the reasons
of such behavior. First-principles computer simulations of
electronic structure and elastic properties of alloys is an
important point making much more effective experimental works on a
purposeful search for ternary alloys with needed elastic properties.
The present work targeted two goals: (i) first-principles
calculations of electronic structure and Young's modulus of binary
Ti$_{1-x}$Nb$_x$ $\beta$-alloy at х=0.07-0.25, and (ii)
determination of the partial substitution of Ti by Zr effect on
electronic structure and elastic moduli of ternary
Ti$_{1-x-y}$Zr$_y$Nb$_x$ $\beta$-alloy.

\section*{Calculation detail}

The high-precision "ab-initio" FLAPW method realized in Wien2k
package \cite{BlahaShwarz01} was used for calculations of elastic
constants $C_{11}$, $C_{12}$ and Young's modulus of Ti$_{1-x}$Nb$_x$
$\beta$-alloy. The exchange-correlation potential was calculated in
generalized gradient approximation (GGA) according to the
Perdew-Burke-Ernzerhof model \cite{BlahaShwarz01}. Atomic sphere
radii both for Ti and Nb atoms were 2.2 a.u.. The calculations were
performed for 2,000 k-points in the first Brillouin zone. Inside the
atomic spheres the wave function was expanded up to lmax=12. The
electronic density and potential inside the spheres were expanded on
the lattice harmonics basis up to Lmax=6. The Rmin$\times$Kmax
parameter which controls the number of APW functions in the basic
set equaled 8.35 that corresponded to 160 APW per atom. The Gmax
value determining the number of plane waves in the inter-sphere
potential expansion equaled 12. These parameters allowed to
calculate the total energy of model ordered structures with accuracy
of 0.001 eV. The calculations were carried out under conditions of a
complete structural optimization considering both homogeneous and
inhomogeneous deformations, i.e. the lattice parameters and atom
locations were optimized. The model volumes were
3а$_{0}\times$3а$_{0}\times$3а$_{0}$ (54 atoms), where а$_0$ is the
BCC lattice parameter of titanium. The elastic moduli were
calculated for ordered structures Ti$_{54-m}$Nb$_{m}$
(m=2,8,10,12,14) which corresponded to  Ti$_{1-x}$Nb$_x$(x=0.037,
0.148, 0.185, 0.222, 0.26) alloys. The large model volume used
provides comparatively high accuracy in calculations of the atomic
structure and deformations caused by substitution of Ti by Nb atoms
in the alloys. The bulk compression, shear, and Young's moduli were
calculated by the Voigt-Reuss-Hill method (VRH) ~\cite{Hill52}. For
ab-initio simulation of elastic properties the approximation of an
elastically isotropic material with $C’$ = $C_{44}$, where
$C’=(C_{11} - C_{12})/2$, was applied. It should be noted that all
calculations realized in the present work correspond to 0 К.

\section*{Results and discussion}

\begin{figure}[t!]
\centering\includegraphics[width=0.7\linewidth]{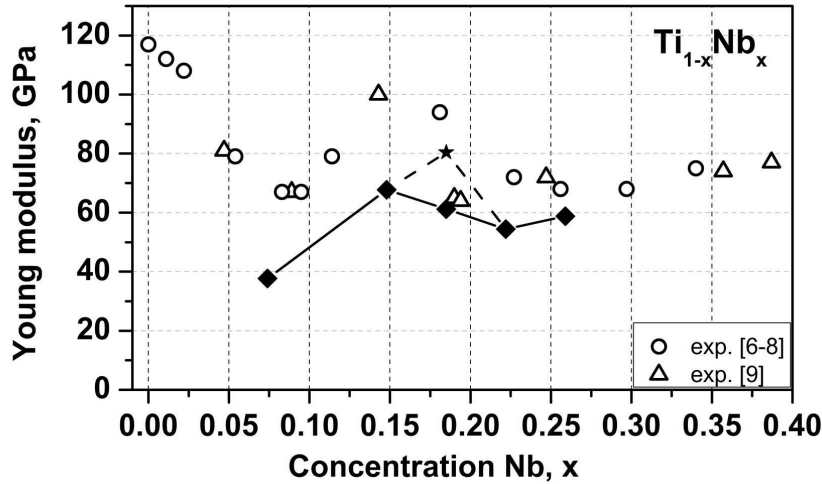}
\caption{Experimental and theoretical concentrational dependences of
Yung modulus of Ti$_{1-x}$Nb$_x$.} \label{chap2:fig1}
\end{figure}

\begin{figure}[h!]
\centering\includegraphics[width=0.7\linewidth]{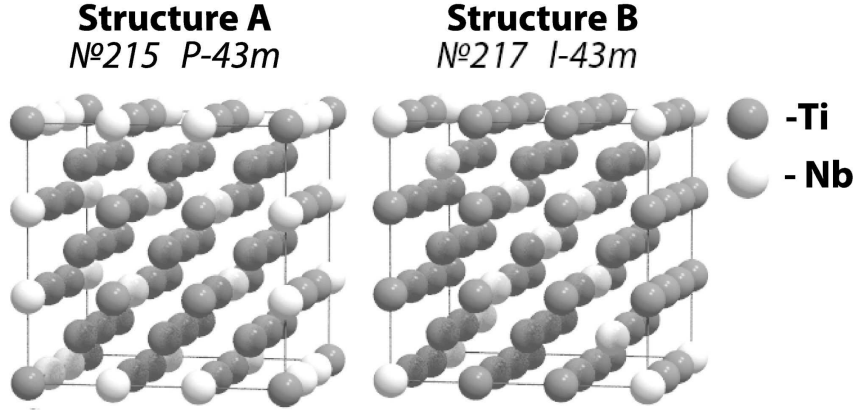} \caption{
Model Structures  Ti$_{44}$Nb$_{10}$.} \label{chap2:fig2}
\end{figure}

For a more adequate simulation of real alloys, Nb atoms in the model
 Ti$_{54-m}$Nb$_{m}$ structures were arranged as uniformly as was possible.
 Each Nb atom did not have any Nb atom adjoining. As mentioned above,
the calculations of total energies for these structures were performed with
a complete structural optimization. The components of elastic
 tensor for the cubic symmetry were found using the CubicElast package (Wien2k
 \cite{BlahaShwarz01}).The structures with tetragonal symmetry were generated from the model cubic
 structures for various Nb contents. Then appropriate deformations with lattice
  parameters changes not higher than 3\% were applied to these structures.
   The total energies were calculated by the FLAPW method for all generated structures.
  Using the total energies of initial and deformed structures,
  the components of elastic tensor for each structure were calculated.
  The calculated equilibrium lattice parameters and elastic constants
  of the $\beta$-titanium and model Ti$_{54-n}$Nb$_n$ (n = 2, 4, 8, 10, 12, 14)
 are listed in Table ~\ref{table1}.
\begin{table}[b!]
\caption{ Lattice constants, elastic constants and Young modulus
(GPa) of Ti (bcc)  and  Ti$_{54-m}$Nb$_m$(m~=~2, 4, 8, 10, 12, 14)
structures modeling binary Ti-Nb alloys.}\label{table1}
\begin{center}
\begin{tabular}{|c|c|c|c|c|c|c|c|}
\hline Alloy &Structure           &$a$, \AA   &$C_{11}$  & $C_{12}$&
$C'$ & $B$  & $E$   \tabularnewline \hline Ti& Ti &3.2553 &79.80
&122.38&-21.29 &108.19  &-68.35 \tabularnewline \hline
 Ti$_{0.962}$Nb$_{0.038}$&Ti$_{52}$Nb$_2$    &3.2500 &104.58  &114.45    &-4.93  &111.16  &-15.02   \tabularnewline \hline
 Ti$_{0.924}$Nb$_{0.076}$&Ti$_{50}$Nb$_4$    &3.2513 &138.31  &122.29    &16.02  &127.63  &37.68    \tabularnewline \hline
 Ti$_{0.852}$Nb$_{0.148}$&Ti$_{46}$Nb$_8$    &3.2662 &145.71  &97.34     &24.19  &113.46  &67.74    \tabularnewline \hline
 \multirow{2}{*}{Ti$_{0.815}$Nb$_{0.185}$} &Ti$_{44}$Nb$_{10}(A)$ &3.2586 &140.57 &97.16      &21.70  &111.63
&61.15    \tabularnewline \cline{2-8}
 &Ti$_{44}$Nb$_{10}(B)$ &3.2715 &152.27 &94.10      &29.09  &113.49  &80.39    \tabularnewline \hline

 Ti$_{0.778}$Nb$_{0.222}$&Ti$_{42}$Nb$_{12}$ &3.2553 &146.67  &108.47    &19.10  &121.20  &54.43    \tabularnewline \hline
 Ti$_{0.741}$Nb$_{0.259}$&Ti$_{40}$Nb$_{14}$ &3.2553 &150.25  &108.84    &20.71  &122.64  &58.81    \tabularnewline
\hline
\end{tabular}
\end{center}
\label{info}
\end{table}
  The concentration dependence of the Young's modulus,
  calculated for the binary $\beta$-alloys, is shown in Fig. ~\ref{chap2:fig1}.
   At low Nb contents negative values of the Young's modulus were obtained,
    that correlates with the   phase field boundaries. At those low Nb
     concentrations initial stresses extensively deform the BCC lattice,
      allowing atoms to occupy energetically more favorable sites proper for
       the HCP phase. At higher Nb contents metastable martensite phases
       energetically more favorable than the $\alpha$ phase form in the real alloys.
        The calculations showed that the concentration dependence of the Young's
         modulus of the binary Ti$_{1-x}$Nb$_x$ alloys is not a monotonic function. This
         dependence has certain peculiar features at Nb contents of х=0.15-0.18 (Fig.~\ref{chap2:fig1}).
      These results correlate well with the experimental data obtained in \cite{Fedotov63,Fedotov64,Fedotov66,Ledbetter80}
      and plotted in the same Figure. Up to date, the Young's modulus increase in the Nb
      concentration range mentioned above has been attributed to the presence of martensite phases.
     However, our results allow to suggest that the increase in Nb concentration
     leads to considerable changes in the electronic structure of the ordered crystal structures, modeling alloy in question.
         It makes a significant influence upon the Young's modulus of the $\beta$ phase. Another
      important factor affecting the Young's modulus is the distribution
      of Nb atoms in the model lattice. This can be demonstrated by
       calculating the Young's modulus for two types of $Ti_{44}Nb_{10}$ structure
        A and B shown in Fig. ~\ref{chap2:fig2}. The A was used earlier for the calculations
    of the Young's modulus  concentration dependence, see Fig. ~\ref{chap2:fig1}.

\begin{figure}[h!]
\centering\includegraphics[width=0.7\linewidth]{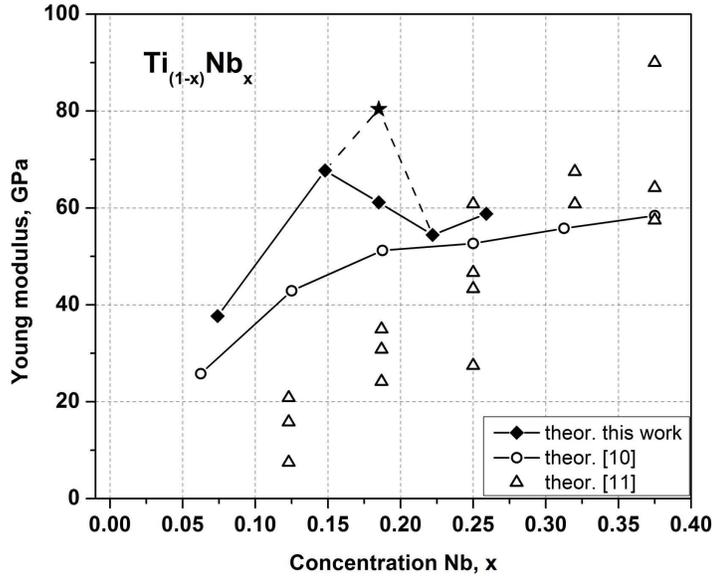}
\caption{Theoretical concentrational dependences of Young modulus of
 Ti$_{1-x}$Nb$_x$ alloy in
\cite{YAO07,Raabe20074475} and this work.}, \label{chap2:fig3}
\end{figure}

    The B was constructed from the A by changing the Nb atom sites.
  In the B each Nb atom has one or several Nb atom neighbors.
  The B structure appeared to be energetically more favorable
  and had higher Young's modulus as compared to the A, see the asterisk in Fig. ~\ref{chap2:fig1}.
  It should be noted that the calculated concentration dependence of the Young's modulus
  better correlates with experimental data if the modulus value for the B
  structure is considered (dashed line in Fig. ~\ref{chap2:fig1}). The calculation
  results allow to formulate two important conclusions. First, the concentration
  dependence of the Young's modulus of the binary single-phase $\beta$-alloy
   is apparently not a monotonic function. So, the peculiar behavior of the experimental
   concentration dependence at certain Nb contents can be explained not only by a multi-phase state.
   Second, the distribution of Nb atoms in ordered structure modeling $\beta$-titanium significantly affects the Young's modulus.
   The structures with a short range ordering, when each Nb atom has one or several Nb neighbors,
   have higher Young's modulus values as compared to the structures without such ordering.
   To clarify the reasons of this effect, a detailed investigation on the nature of the
   chemical bonds in Ti-Nb system and the mechanism of the $\beta$-phase stabilization by Nb is needed.
   Unfortunately, these problems are beyond the scope of present paper. The discrepancy
   between the calculated and experimental Young's modulus values can be explained
   by using polycrystalline and multi-phase samples in the experimental works.
   Our results significantly differ from the theoretical results obtained in other works \cite{YAO07,Raabe20074475},
   see Fig. ~\ref{chap2:fig3} for comparison.
   Our theoretical values have the best correlation
   with experimental data, see Fig. ~\ref{chap2:fig1}. The peak on the theoretical concentration
   dependence at х=0.185 is a peculiar feature of our results. The better correlation with
   experiment is reached for the absolute values of the Young's modulus as well. The values obtained in \cite{YAO07,Raabe20074475}
   are significantly lower than both ours and experimental ones. The main reason of the better
   correlation of our results with experiment is possibly a larger model volume used (54 atoms),
   while in the works mentioned above model volume comprised 16 atoms (2а$_{0}\times$2а$_{0}\times$2а$_{0}$,
    where а$_0$ is the BCC lattice parameter). In \cite{YAO07} the FLAPW method for
    the Young's modulus calculations was applied as well. The elastic properties were
    simulated by the Voigt-Reuss-Hill approximation (VRH) ~\cite{Hill52} for polycrystals.
    In \cite{Raabe20074475} the Young's modulus was calculated by the
    pseudo-potential method with the VASP package \cite{PhysRevB.54.11169}
    in the elastically isotropic approximation. The Young's modulus of several
    model structures for fixed Nb contents but different atom distributions were
    also calculated in \cite{PhysRevB.54.11169}, the results are included in fig. ~\ref{chap2:fig3}.
    Unfortunately, the authors did not describe the principles of structure selection used.
    The insufficient model volumes used in the two earlier works, in our opinion, is
    one of the reasons of a considerable discrepancy between the theoretical and experimental values.
    The three-fold larger model volume used in the present work apparently allowed a more
    adequate modeling the real atomic structure, and a more accurate calculation of
    deformation effects caused by substitution of Ti by Nb atoms. The concentration dependence
    of the Young's modulus of the binary Ti-Nb $\beta$-alloy makes possible modeling
    the elastic properties of ternary alloys. Our aim is to predict the compositions
    providing single-phase state and Young's modulus values lower as compared to
    the binary alloy. As mentioned in the Introduction, Zr as the third element is promising.
    Model ordered structures of ternary alloy were formed from the A and B structures shown in Fig. ~\ref{chap2:fig2}.
    The Nb content in these structures was constant, х=0.185. The Young's modulus calculated
    for this concentration was quite high, Table ~\ref{table1}. So, it is of interest to substitute
    some Ti by Zr atoms in proportions that would result both in a further reduction
    of the Young's modulus as compared to the binary alloy, and in single-phase state.
    Two model structures of the A type with differing Zr content, Ti$_{44-m}$Zr$_{m}$Nb$_{10}$(m~=~16,
28), were chosen. These structures correspond to
{Ti$_{0.519}$Zr$_{0.296}$Nb$_{0.185}$} and
{Ti$_{0.296}$Zr$_{0.519}$Nb$_{0.185}$} alloys. As earlier for the binary alloys,
 the $\beta$-type structures with a peculiar Nb distribution were constructed from the A structures.
 In the B structures Nb atoms had one or several Nb neighbors. As for the binary alloys,
 the calculations showed high energetic favorability of the $\beta$-type structures.
 The calculations of the atomic structure and elastic constants were fulfilled
 for these four model structures. The results are listed in Table \ref{table2}.
\begin{table}[b!]
\caption{ Lattice constants, elastic constants and Young modulus
(GPa) of Ti$_{44-m}$Zr$_{m}$Nb$_{10}$(m~=~16, 28) structures
modeling  triple  Ti-Zr-Nb alloys.}\label{table2}
\begin{center}
\begin{tabular}{|c|c|c|c|c|c|c|c|}
\hline Alloy &Structure           &$a$, \AA   &$C_{11}$  & $C_{12}$&
$C'$ & $B$  & $E$   \tabularnewline \hline
 \multirow{2}{*}{Ti$_{0.519}$Zr$_{0.296}$Nb$_{0.185}$} &Ti$_{28}$Zr$_{16}$Nb$_{10}$(A)&3.3520 &130.00  &100.00    &15.00  &110.00  &43.04    \tabularnewline
 \cline{2-8}
 &Ti$_{28}$Zr$_{16}$Nb$_{10}$(B)&3.3820 &144.22  &94.66    &24.78  &111.18  &69.19   \tabularnewline
 \hline

 \multirow{2}{*}{Ti$_{0.296}$Zr$_{0.519}$Nb$_{0.185}$}&Ti$_{16}$Zr$_{28}$Nb$_{10}$(A)&3.4253 &129.01 &94.67      &17.5  &105.67  &49.75    \tabularnewline
 \cline{2-8}
 &Ti$_{16}$Zr$_{28}$Nb$_{10}$(B)&3.4315 &137.12  &89.70     &23.71  &105.5  &66.18    \tabularnewline
 \hline

\hline
\end{tabular}
\end{center}
\label{info2}
\end{table}
\begin{figure}[t!]
\centering\includegraphics[width=0.7\linewidth]{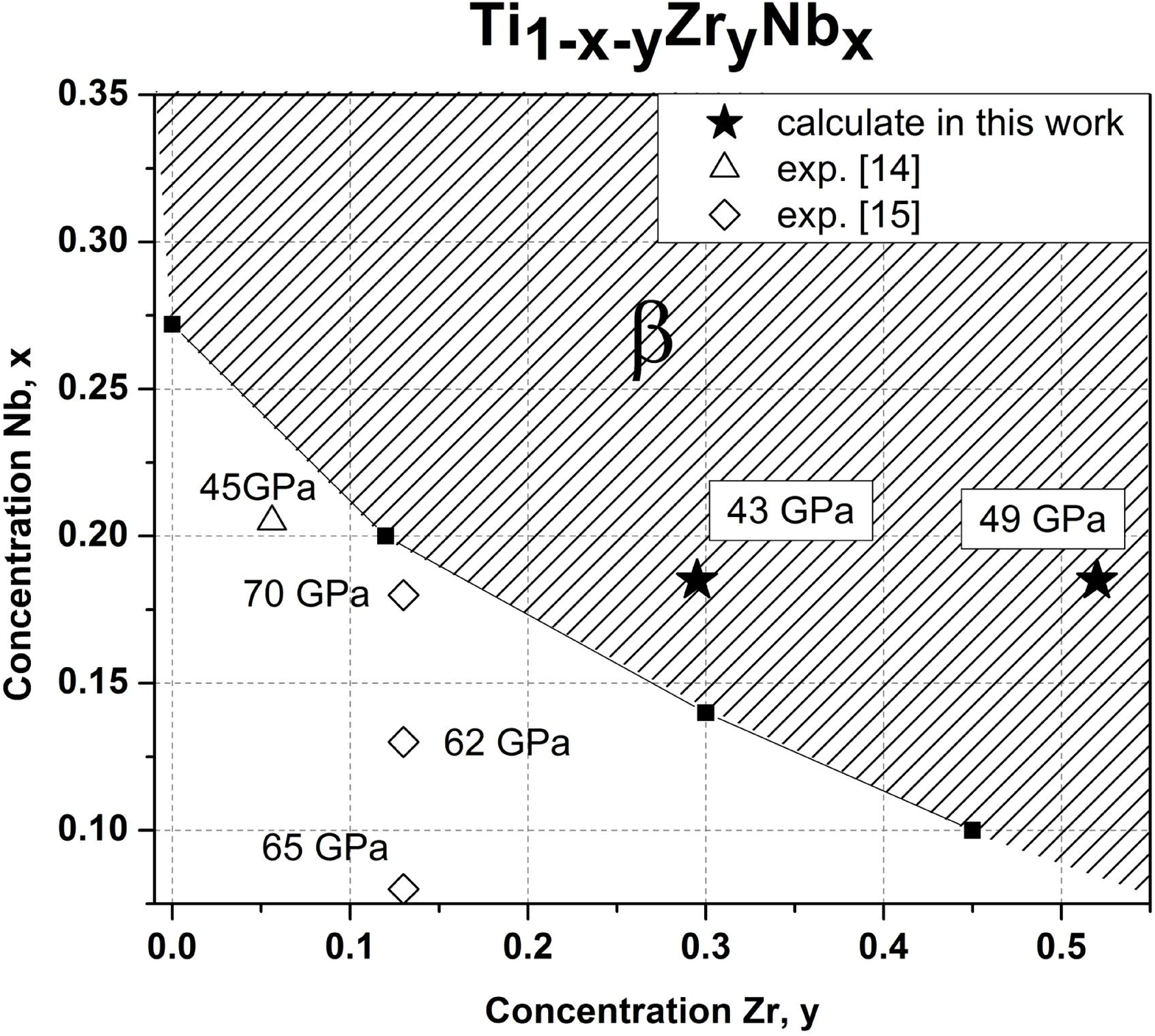}
\caption{Region of existence of single phase and multi phase triple
$\beta$-phases Ti-Zr-Nb} \label{chap2:fig5}
\end{figure}
  The substitution
 of Ti by 30\% Zr atoms in the Ti$_{44}$Nb$_{10}$(A) structure led to a reduction of the
  Young's modulus from 61.15 to 43.04 GPa in the Ti$_{28}$Zr$_{16}$Nb$_{10}$(A) structure.
  Further increase in Zr content up to 52 at.\% resulted in the Young's modulus
  of 49.75 GPa in the Ti$_{16}$Zr$_{28}$Nb$_{10}$(A)) structure.
  Therefore, our results predict a significant reduction of the Young's modulus
  in the ternary alloys as compared to binary ones. However, the first-principles
  calculations can not predict whether the ternary alloys will be single-phase or not.
  To clarify this question we used experimental results of \cite{AbdelHady20071000}.
  In this work the influence of Zr on the phase composition of ternary Ti-Zr-Nb alloys
   in a wide concentration range was investigated. The authors showed that partial
   substitution of Ti by Zr in binary Ti-Nb alloy led to the $\beta$-phase stabilization.
   The samples annealed at 1173 К were investigated. Based on the results of \cite{AbdelHady20071000},
   we plotted the dependence of the lowest Nb content providing single-phase state on Zr
   concentration, see Fig. ~\ref{chap2:fig5}. This allowed to estimate the single-phase  $\beta$
   field boundaries for the ternary alloys. The alloys modeled in the present work are designated
    by asterisk. The Young's modulus values calculated for the A-type structures are indicated.
   Apparently, the search for compositions providing low Young's modulus values is most
   effective in the hatched area at moderate Nb contents. Experimental results of \cite{Brailovski2011643,Schneidera05}
   are also included for comparison. The samples studied in those works comprised
   several phases, that resulted in comparatively high Young's modulus values.

\section*{Conclusions}

 The concentration dependence of the Young's modulus of
several ordered Ti$_{54-m}$Nb$_{m}$ structures, modeling binary
Ti$_{1-x}$Nb$_x$ alloys, was calculated by first-principles methods.
It was shown that the features of this dependence at 0.15-0.18 at.\%
Nb can be attributed to the types of Nb atoms distribution. A
detailed comparison of the theoretical results with experimental
data was fulfilled. The Young's modulus values were calculated for a
set of ordered structures with different Nb atoms distribution,
modeling ternary alloys Ti$_{0.518}$ Zr$_{0.297}$Nb$_{0.185}$ and
Ti$_{0.297}$Zr$_{0.518}$Nb$_{0.185}$. The results of these
calculations allowed us to suggest the concentration region for
single-phase ternary $\beta$-phase alloys possessing low values of
Young's modulus.

\bibliographystyle{unsrt}
\bibliography{literature}

\end{document}